\documentstyle[12pt]{article}

\newcommand{\ie}{{\it i.e.}}
\newcommand{\eg}{{\it e.g.}}

\newcommand{\lsim}{\buildrel < \over {_\sim}}

\newcommand{\ccbar}{c\bar{c}}

\newcommand{\jp}{J/\psi}
\newcommand{\as}{$\alpha_s$}

\newcommand{\PLB}[3]{\mbox{}{\it Phys. Lett.} {\bf B{#1}}, {#2} ({#3})}
\newcommand{\NPB}[3]{\mbox{}{\it Nucl. Phys.} {\bf B{#1}}, {#2} ({#3})}
\newcommand{\PRL}[3]{\mbox{}{\it Phys. Rev. Lett.} {\bf {#1}}, {#2} ({#3})}
\newcommand{\PRD}[3]{\mbox{}{\it Phys. Rev.} {\bf D{#1}}, {#2} ({#3})}
\newcommand{\ZPC}[3]{\mbox{}{\it Z. Phys.} {\bf C{#1}}, {#2} ({#3})}

\newcommand{\etal}{{\em et al.}}
\begin{document}

\thispagestyle{empty}
\begin{flushright}
   \vbox{\baselineskip 12.5pt plus 1pt minus 1pt
         SLAC-PUB-6665 \\
         September 1994 \\
         (T/E)
             }
\end{flushright}

\renewcommand{\thefootnote}{\fnsymbol{footnote}}
\bigskip
\begin{center}
{\Large  \bf Polarization
of Charmonium in $\pi N$ Collisions}

\vskip 1\baselineskip

{Wai-Keung Tang\footnote{Work supported
by Department of Energy contract DE--AC03--76SF00515.}
\footnote{Presented
at The Eleventh International Symposium on High Energy Spin Physics and
the Eighth International Symposium on Polarization Phenomena in Nuclear
Physics, Bloomington, In.,
September 15-22, 1994.}
} \\
{\footnotesize \em Stanford Linear Accelerator Center,
Stanford University, Stanford, CA 94309}

\end{center}

\medskip

\renewcommand{\thefootnote}{\arabic{footnote}}
\addtocounter{footnote}{-2}

\vbox{\footnotesize
\noindent
{\bf Abstract.}
Measurements of the polarization of $\jp$ produced in pion-nucleus
collisions are in disagreement with
leading twist QCD prediction where  $\jp$
is observed to have negligible polarization whereas theory predicts
substantial polarization.
We argue that this discrepancy cannot be due to
poorly known
structure functions nor the  relative production
rates of $\jp$ and $\chi_J$.
The disagreement between theory and experiment suggests important
higher twist corrections, as has earlier been surmised
from the anomalous non-factorized nuclear
$A$-dependence of the $\jp$ cross section.
}

\pagestyle{empty}

\vspace{6 mm}

\begin{center}
{\large \bf INTRODUCTION}
\end{center}


One of the most sensitive tests of the QCD mechanisms for the
production of heavy quarkonium is the polarization of the $\jp$  in
hadron collisions. In fact, there are serious disagreements between
leading twist QCD
prediction \cite{BargerPRD31} and experimental data
\cite{Clark}
on the production cross section
of `direct' $\jp$ and
$\chi_1.$ We would like to advocate that polarization of $\jp$ provides
strong constraints on the production mechanisms of $\jp$ and thus can
pinpoint the origin of these disagreements.

In this paper we will present results on the
theoretical calculation of the polarization of $\jp$ in $\pi N$
collisions. The detailed analysis will be published in a later
paper\cite{mhbt}. We found that the polarization of $\jp$ provides
important constraints on the nature of the production mechanisms and
urge that polarization measurement of $\jp$
should be included in the design of
future charm production experiment.

\vspace{6 mm}

\begin{center}
{\large \bf PRODUCTION RATES OF $\psi$ AND $\chi_J$ STATES}
\end{center}


In leading twist QCD,
the production of the
$\jp$ at low transverse momentum occurs both `directly'
from the gluon fusion subprocess $gg \to \jp+g$
and indirectly
via the production and decay of $\chi_1$ and $\chi_2$ states.
These states
have
sizable decay branching fractions $\chi_{1,2} \to \jp+\gamma$
of 27\% and 13\%,
respectively.

In this model, we assume that
the non-perturbative physics, which is described by the wave function
at the origin in cases of production of $\jp$ and $\psi'$, is
separable from the perturbative hard subprocess, $\ie,$ factorization holds. As
the wave function at the
origin can be related to the leptonic decay amplitude, the ratio of
$\psi'$ to direct $\jp$ production can be expressed in terms of the
ratio of their leptonic decay width. More precisely, taking into account
of the phase space factor,
\begin{equation}
\frac{\sigma(\psi')}{\sigma_{dir}(\jp)} \simeq \frac{\Gamma(\psi'\to e^+e^-)}
  {\Gamma(\jp\to e^+e^-)} \frac{M_{\jp}^3}{M_{\psi'}^3} \simeq 0.24 \pm 0.03
  \label{psiprime}
\end{equation}
where $\sigma_{dir}(\jp)$ is the cross section for direct production of the
$\jp$. The
ratio (\ref{psiprime}) should hold for all beams and targets, independently of
the size of the higher twist corrections in producing the point-like $\ccbar$
state. The energy should be large enough for the bound state to form outside
the
target. The available data is indeed compatible with (\ref{psiprime}). In
particular, the E705 values \cite{AntoniazziPRL70} with different
projectiles are all consistent with $0.24$.


The anomalous nuclear target $A$-dependence observed for the $\jp$
is also seen for the $\psi'$ \cite{AldeJP}, so that the ratio (\ref{psiprime})
is indeed independent of $A$. Therefore, at high energies, the
quarkonium bound state forms long after the production of the $\ccbar$ pair
and the formation process is well described by the non-relativistic
wavefunction at the origin.




In leading twist and to
leading order in \as,  $\jp$  production can be
computed from the convolution of hard subprocess cross section
$gg\rightarrow \jp g$, $gg\rightarrow \chi_{j}$, {\em etc.},
with the parton
distribution functions in the beam and target. Higher order corrections in \as,
and relativistic
corrections to the charmonium bound states, are unlikely to change our
qualitative conclusions at moderate $x_{F}$. Contributions from direct $\jp$
production, as well as
from indirect production via $\chi_1$ and $\chi_2$ decays, will be included.
Due
to
the small branching fraction $\chi_0 \to \jp+\gamma$ of 0.7\%, the contribution
from $\chi_0$ to $\jp$ production is expected (and observed) to be negligible.
Decays from the radially excited $2^3S_1$  state, $\psi' \to \jp + X$,
contribute
to  the total $\jp$ rate at the few per cent level and will be ignored here.

In Table 1 we compare the $\chi_2$ production
cross section, and the relative rates of direct $\jp$ and $\chi_1$ production,
with the data of E705 and WA11 on $\pi^-N$ collisions at $E_{lab}=300$ GeV
and 185 GeV \cite{AntoniazziPRL70}.
The $\chi_2$ production rate in QCD agrees with the data within a
`$K$-factor' of order $2$ to $3$. This is within the theoretical uncertainties
arising from the $\jp$ and $\chi$ wavefunctions, higher order corrections,
structure functions, and the
renormalization scale. A similar factor is found between the lowest-order
QCD calculation and the data on lepton pair production \cite{BadierDY}
On the
other hand, Table 1 shows a considerable discrepancy between the calculated and
measured relative production rates of direct $\jp$ and $\chi_1$, compared to
$\chi_2$ production. A {\em priori} we would expect the $K$-factors to be
roughly
similar for all three processes.
We conclude that leading twist QCD appears to be in conflict with the data on
direct $\jp$ and $\chi_1$ production. Although in Table 1 we have only compared
our
calculation with the E705 and WA11
$\pi^- N$ data, this comparison is representative of
the overall situation (for a recent comprehensive review see \cite{Schuler}).
\begin{table}[tb]
\begin{center}
\begin{tabular}{|c|c|c|c|}
  \hline
  & $\sigma(\chi_2)$ [nb] & $\sigma_{dir}(\jp)/\sigma(\chi_2)$
    & $\sigma(\chi_1)/\sigma(\chi_2)$ \\ \hline
  Experiment & $188 \pm 30\pm 21$  & $0.54 \pm 0.11\pm 0.10$ & $0.70 \pm
  0.15\pm 0.12$
   \\ \hline
  Theory     & 72 & 0.19 & 0.069 \\ \hline
\end{tabular}
\end {center}
\caption{Production cross sections for $\chi_1$, $\chi_2$ and directly produced
$\jp$ in $\pi^- N$ collisions. The data from Ref.
\protect\cite{AntoniazziPRL70,E705}
include measurements at 185 and 300 GeV. The theoretical
calculation is at 300 GeV.}
\end{table}

\vspace{6 mm}

\begin{center}
{\large \bf POLARIZATION OF THE $\jp$
}
\end{center}


The polarization of the $\jp$ is determined by the angular distribution of its
decay muons in the $\jp$ rest frame.
 The
angular distribution of massless muons, integrated over the azimuthal angle,
has
the form
\begin{equation} \frac{d\sigma}{d\cos\theta} \propto 1 + \lambda \cos^2
\theta \label{lambda}
\end{equation}
where we take $\theta$ to be the angle between the $\mu^+$ and the projectile
direction (\ie, we use the Gottfried--Jackson frame). The parameter $\lambda$
can
be calculated from the $\ccbar$ production amplitude and the electric dipole
approximation of radiative $\chi$ decays.


In Fig.~1a we show the predicted value (solid curve) of the parameter $\lambda$
of Eq. (\ref{lambda})
in the GJ-frame as a function of $x_F$, separately for the direct $\jp$ and the
$\chi_{1,2} \to
\jp+\gamma$ processes. Direct $\jp$ production gives $\lambda \simeq 0.25$,
whereas the production via $\chi_1$ and $\chi_2$
result in $\lambda \simeq -0.15$ and 1 respectively.
Smearing of the beam
parton's transverse momentum distribution by a Gaussian function
$\exp \left[ -(k_\perp/500 \; {\rm MeV})^2 \right] $ (dashed curve)
has no significant
effect in $\lambda$ except for the production via $\chi_2$ which
brings
$\lambda$ down to $\lambda \simeq 0.85$.
The $\lambda(x_F)$-distribution obtained when both the direct
and indirect
$\jp$ production processes are taken into account is shown in Fig.~1b and
is
compared with the Chicago--Iowa--Princeton
 \cite{Biino}
and E537 data \cite{Akerlof}
 for 252 GeV $\pi W$
collisions and 150 GeV $\pi^- W$ collisions respectively.
 Our QCD calculation gives $\lambda \simeq 0.5$ for $x_F \lsim
0.6$, significantly different from the measured value $\lambda \simeq 0$.

\begin{figure}[tb]
\vspace{2.7 in}
\caption{CIP($\bullet$) and E537 ($\circ$) data compared with theoretical
prediction.
}
\label{polarizat}
\end{figure}

The discrepancies between the calculated and measured values of $\lambda$
is one further indication that the standard leading twist processes
considered here are not adequate for explaining charmonium production.
The $\jp$ polarization is particularly sensitive to the production
mechanisms and allows
 us to make further conclusions on
the origin of the disagreements, including the above discrepancies in
the relative production cross sections of $\jp$, $\chi_{1}$ and
$\chi_{2}$. If these discrepancies arise from an incorrect
relative normalization of the various subprocess contributions (\eg, due to
higher order effects), then we would expect the $\jp$ polarization to agree
with data when the relative rates of the subprocesses are adjusted according to
the measured cross sections of direct $\jp$, $\chi_1$ and $\chi_2$
production
 The lower curve in Fig.~1b shows the effect of multiplying the partial
$\jp$ cross sections with the required $K$-factors.
The smearing effect is insignificant
as shown by the dashed curve.
The $\lambda$ parameter is
still predicted incorrectly over most of the $x_F$ range.

A similar conclusion is reached (within somewhat larger experimental errors) if
we compare our calculated value for the polarization of direct $\jp$
production,
shown in Fig.~1a,
with the measured value of $\lambda$ for $\psi'$ production. In analogy
to Eq.~(\ref{psiprime}), the $\psi'$ polarization data
should agree with the polarization of directly produced
$\jp$'s, regardless of the production mechanism.  Based on the angular
distribution of the muons from $\psi' \to \mu^+\mu^-$ decays in 253 GeV
$\pi^-W$
collisions, Ref. \cite{Heinrich} quotes $\lambda_{\psi'} = 0.02 \pm 0.14$ for
$x_F>0.25$, appreciably smaller than our QCD values for direct $\jp$'s in
Fig.~1a.

\vspace{6 mm}

\begin{center}
{\large \bf DISCUSSION
}
\end{center}


We have seen that the $\jp$ and $\chi_1$ hadroproduction cross sections in
leading twist QCD are at considerable variance with the data, whereas the
$\chi_2$
cross section agrees with measurements within a reasonable $K$-factor of
$2$ to $3$. On
the other hand,
the relative rate of $\psi'$ and direct $\jp$ production (Eq.~\ref{psiprime}),
which at high energies
should be independent of the production mechanism, is in agreement with
experiment.
It is therefore improbable that the treatment of the $\ccbar$ binding
should require large corrections.

In a leading twist description, an incorrect normalization of the
charmonium production cross sections can arise from large higher order
corrections or uncertainties in the parton distributions\cite{Schuler}.
Taking into account that the normalization may be wrong by as much as a factor
of 10 and that even such a $K-$factor does not explain the polarization
data of $\jp$,
a more likely
explanation may be that
there are important
higher-twist contributions to the production of the $\jp$ and $\chi_1$
as suggested in large $x_F$ case \cite{BHMT,HVS}.

Further theoretical work is needed to establish that the data on direct $\jp$
and $\chi_1$ production indeed can be described from  higher twist
mechanisms.
 Experimentally, it is important to check whether the
$\jp$'s produced indirectly via $\chi_2$ decay are transversely polarized. This
would show that $\chi_2$ production is dominantly leading twist, as we have
argued. Thus, the polarization of $\jp$ production from different
channels provides a very sensitive discriminant of different
production mechanisms.





\vspace{6 mm}
\begin{center}
{\large \bf REFERENCES}
\end{center}

\begin{enumerate}
{\footnotesize
\item M. V\"anttinen, P. Hoyer, S. J. Brodsky and W.-K. Tang,
preprint SLAC-PUB-6637 and HU-TFT-94-29.
\item V. Barger and A. D. Martin, \PRD{31}{1051}{1985}.
\item A. G. Clark, \etal, \NPB{142}{29}{1978};
 R806: C. Kourkoumelis \etal, \PLB{81}{405}{1979};
 WA11: Y. Lemoigne, \etal, \PLB{113}{509}{1982};
 E673: S. R. Hahn, \etal, \PRD{30}{671}{1984}; D. A. Bauer,
 \etal, \PRL{54}{753}{1985};
 F. Binon, \etal, \NPB{239}{311}{1984}.

\item E705: L. Antoniazzi, \etal, \PRL{70}{383}{1993}

\item E705: L. Antoniazzi, \etal, \PRD{46}{4828}{1992}

\item G. A. Schuler, preprint CERN-TH.7170/94.

\item C. Biino, \etal, \PRL{58}{2523}{1987}.

\item E537: C. Akerlof, \etal, \PRD{48}{5067}{1993}.

\item E772: D. M. Alde, \etal, \PRL{66}{133}{1991}.

\item J. Badier, \etal, \ZPC{18}{281}{1983}.
      J. S. Conway, \etal, \PRD{39}{92}{1989}.


\item J. G. Heinrich, \etal, \PRD{44}{1909}{1991}.



\item S. J. Brodsky, P. Hoyer, A. H. Mueller and W.-K. Tang,
\NPB{369}{519}{1992}.

\item P. Hoyer, M. V\"anttinen and U. Sukhatme,
\PLB{246}{217}{1990}.}

\end{enumerate}

\end{document}